\documentclass[prd,twocolumn,preprintnumbers,showpacs,showkeys,superscriptaddress]{revtex4}
\usepackage{amsmath}
\usepackage{graphicx}

\newcommand\vek[1]{{\boldsymbol{#1}}}
\newcommand\adj[1]{\overline{#1}}
\newcommand\cc[1]{#1^{\mathcal C}}
\newcommand\gr[1]{\mathrm{#1}}

\begin{document}
\preprint{YITP-09-44, KUNS-2217}

\title{Two-color quark matter:  $\gr{U(1)_A}$ restoration,
 superfluidity, and quarkyonic phase}
\author{Tom\'{a}\v{s} Brauner%
\footnote{On leave from Department of Theoretical Physics, Nuclear
 Physics Institute ASCR, CZ-25068 \v Re\v z, Czech Republic}}
\email{brauner@ujf.cas.cz}
\affiliation{Institut f\"ur Theoretische Physik, Goethe-Universit\"at,
Max-von-Laue-Stra\ss e 1, D-60438 Frankfurt am Main, Germany}
\author{Kenji Fukushima}
\affiliation{Yukawa Institute for Theoretical Physics,
 Kyoto University, Kyoto 606-8502, Japan}
\author{Yoshimasa Hidaka}
\affiliation{Department of Physics, Kyoto University,
 Kyoto 606-8502, Japan}

\begin{abstract}
 We discuss the phase structure of quantum chromodynamics (QCD) with
 two colors and two flavors of light quarks.  This is motivated by the
 increasing interest in the QCD phase diagram as follows:  (1) The QCD
 critical point search has been under intensive dispute and its
 location and existence suffer from uncertainty of effective
 $\gr{U(1)_A}$ symmetry restoration.  (2) A new phase called
 quarkyonic matter is drawing theoretical and experimental attention
 but it is not clear whether it can coexist with diquark condensation.
 We point out that two-color QCD is nontrivial enough to contain
 essential ingredients for (1) and (2) both, and most importantly, is
 a system without the sign problem in numerical simulations on the
 lattice.  We adopt the two-flavor Nambu--Jona-Lasinio model extended
 with the two-color Polyakov loop and make quantitative predictions
 which can be tested by lattice simulations.
\end{abstract}

\maketitle


\section{Introduction}

Understanding the phase structure of quantum chromodynamics (QCD) is
one of the key issues in current high-energy physics. Thorough
phenomenological knowledge of properties of the hadron spectrum as
well as nuclear matter is now being complemented by increasingly
precise first-principle numerical studies of QCD at nonzero
temperature.  However, the application of lattice techniques to matter
at high baryon chemical potential $\mu_B$ remains a major challenge
due to the infamous sign problem~\cite{Muroya:2003qs}.  The
difficulties encountered in simulations of QCD triggered interest in
similar theories which are free of the sign problem.  These include
simulations at imaginary chemical
potential~\cite{Alford:1998sd,deForcrand:2002ci,D'Elia:2002gd,
deForcrand:2003hx,D'Elia:2004at,Chen:2004tb,Azcoiti:2005tv,deForcrand:2006pv,Wu:2006su,
D'Elia:2007ke,deForcrand:2008vr,D'Elia:2009tm}, QCD at nonzero isospin
density~\cite{Son:2000xc,Kogut:2002zg,Kogut:2004zg}, a QCD-like theory
with adjoint
quarks~\cite{Kogut:2000ek,Splittorff:2000mm,Hands:2000ei,Hands:2001ee},
and two-color
QCD~\cite{Nakamura:1984uz,Kogut:1999iv,Kogut:2001na,Muroya:2002ry,%
Kogut:2003ju,Skullerud:2003yc,Giudice:2004se,Chandrasekharan:2006tz,Hands:2006ve,%
Cea:2007vt,Hands:2007uc,Hands:2008ha,Lombardo:2008vc}.  The latter, two-color QCD, will be the subject of the
present article.

Two-color QCD differs in several aspects from the world we live in.
The most notable difference perhaps is that the colorless baryons are
formed from two quarks, and hence are bosons.  Dense matter is then
not realized as an interacting Fermi sea of nucleons, but rather as a
Bose gas of diquarks which undergoes Bose--Einstein condensation (BEC)
at sufficiently low temperature.  Therefore the ground state of cold
and dense two-color quark matter forms a
superfluid~\cite{Rapp:1998zu}.  Another noteworthy feature of
two-color QCD, stemming from the fact that the $\gr{SU(2)}$ gauge
group has only pseudo-real representations, is the Pauli--G\"ursey
symmetry~\cite{Smilga:1995tb} connecting quarks with antiquarks.  As a
consequence, the spectrum of Nambu--Goldstone (NG) bosons of the
spontaneously broken chiral symmetry contains diquark states in
addition to the pseudoscalar mesons.  The presence of light particles
carrying baryon number (i.e.~baryonic pions) is quite peculiar to
two-color QCD (and QCD with adjoint quarks) and crucial for
understanding the phase structure at $T\neq0$ and $\mu_B\neq0$ by means of the
chiral Lagrangian
approach~\cite{Kogut:2000ek,Splittorff:2000mm,Splittorff:2001fy,
Splittorff:2002xn,Kanazawa:2009ks}. The model-independent
chiral Lagrangian arguments have been
complemented by investigations in various models such as the linear
sigma model~\cite{Lenaghan:2001sd,Wirstam:2002be}, the random matrix
theory~\cite{Vanderheyden:2001gx,Klein:2004hv,Shinno:2009jw}, the strong-coupling
expansion~\cite{Dagotto:1986gw,Chandrasekharan:2006tz,%
Chandrasekharan:2006cd,Nishida:2003uj,Fukushima:2008su}.  The
Nambu--Jona-Lasinio (NJL) model was first applied to two-color QCD in
Ref.~\cite{Ratti:2004ra}.

In fact there has been a tight communication between a number of
effective model studies and the Monte-Carlo simulations on the
lattice.  Recent progress in this direction led to first attempts to
probe BEC of diquarks and the region of moderate baryon
density~\cite{Hands:2006ve,Hands:2008ha}.
Thus, based on the knowledge achieved in these preceding papers, the
present paper aims to make a proposal to use the two-color QCD model
as a controllable test setting to clarify the following controversial
issues on the real-QCD phase diagram.

\vspace{2mm}
\noindent
--- \textbf{In-medium $\gr{U(1)_A}$ Symmetry Restoration} ---\\
The QCD Lagrangian has global $\gr{U(1)_A}$ symmetry at the classical level.
The quantum anomaly, however, leads to a nonconserving contribution to
the axial current, which breaks $\gr{U(1)_A}$ symmetry explicitly.
From the point of view of quantum field theory the anomaly
comes from highly ultraviolet modes, and thus, the anomaly should be
insensitive to any infrared scales such as $T$, $\mu_B$, and the quark
mass $m_0$.  In this sense the $\gr{U(1)_A}$ anomaly is never restored
at any $T$ nor $\mu_B$.  In the effective model description the
$\gr{U(1)_A}$ anomaly manifests itself in the form of a
$\gr{U(1)_A}$-breaking interaction which is microscopically induced by
instantons~\cite{'tHooft:1976up}.  Because instantons are suppressed
at high $T$ or $\mu_B$~\cite{Gross:1980br,Schafer:2002ty}, the
$\gr{U(1)_A}$-breaking interaction is anticipated to weaken in a
medium, leading to ``effective restoration'' of the $\gr{U(1)_A}$
symmetry~\cite{Pisarski:1983ms,Shuryak:1993ee,Fukushima:2001hr,%
Costa:2004db,Costa:2008dp}.  In fact, a quantitative estimate of the
reduction of the $\gr{U(1)_A}$ effect is crucial for locating the QCD
critical point in the $\mu_B-T$ phase
diagram~\cite{Chandrasekharan:2007up,Fukushima:2008wg,Chen:2009gv}.
The numerical study by the lattice Monte-Carlo simulation is in
principle possible in two-color QCD. The clear signal for
$\gr{U(1)_A}$ restoration is degeneracy in the spectra of mesons
connected by a $\gr{U(1)_A}$ rotation~\cite{Shuryak:1993ee}.  That is,
in the two-flavor case, the $\gr{U(1)_A}$ partners are
\begin{equation*}
 \begin{split}
 &\left\{ \begin{array}{l}
  \text{Scalar-isoscalar ($\sigma$) meson,} \\
  \text{Pseudoscalar-isoscalar ($\eta_0$) meson,}
 \end{array} \right. \\
 &\left\{ \begin{array}{l}
  \text{Scalar-isovector ($\vec{a}_0$) meson,} \\
  \text{Pseudoscalar-isovector ($\vec{\pi}$) meson.}
 \end{array} \right.
 \end{split}
\end{equation*}
The masses of these multiplets become degenerate when the chiral symmetry is
also restored.

Since the $\sigma$ meson involves the so-called disconnected diagrams,
the lattice simulation is too noisy to see the degeneracy with
$\eta_0$ (flavor-singlet $\eta$) which is also in a noisy channel.  It
is, however, feasible to check the degeneracy between $\vec{a}_0$ and
$\vec{\pi}$ in the finite-$T$ and finite-$\mu_B$ lattice simulation of
two-color QCD.\ \ In this paper we will give a quantitative guide from
an effective model study.

\begin{figure}
\includegraphics[width=\columnwidth]{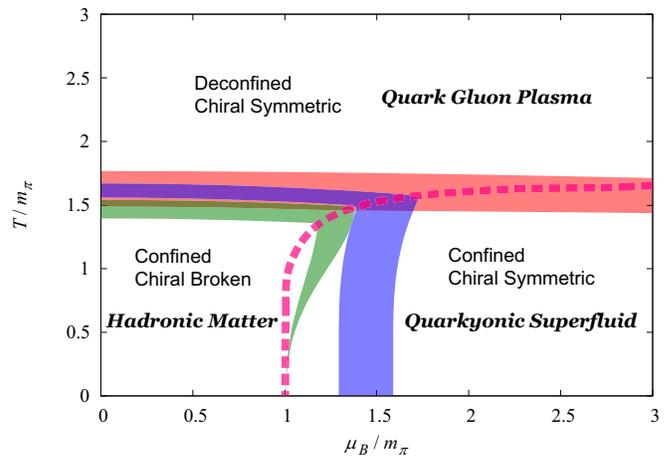}
\caption{Phase diagram of two-color QCD with two light quark flavors.
  The red and blue bands represent the regions in which the Polyakov loop and the normalized
  chiral condensate, respectively, take a value ranging from 0.4 to
  0.6.  The red dashed line extending from the bottom to the right
  shows the onset of diquark condensation.  The green band surrounding
  the left-bottom corner indicates the region where $n_B$ normalized
  by the ``Stefan--Boltzmann'' (see Sec.~\ref{sec:phase} for detailed
  explanations) value ranges from 0.4 to 0.6.}
\label{fig:phase}
\end{figure}

\vspace{2mm}
\noindent
--- \textbf{Superfluidity and Quarkyonic Matter} ---\\
A new state of matter at high density has been recognized and it is
now referred to as
\textit{quarkyonic matter}~\cite{McLerran:2007qj,Hidaka:2008yy}.
Because the definition of quarkyonic matter is clear only in the large
$N_c$ limit, in which color non-singlet interactions are subleading,
there seem to be some confusing arguments like that quarkyonic matter
overwhelms color superconductivity.  This is not quite true especially
when $N_c$ is finite.  Intuitively, quarkyonic matter is characterized
by the properties that the thermodynamic quantities (pressure, baryon
density, etc.) should be almost saturated by the degenerated Fermi liquid of
quarks and the collective excitations on top of the Fermi surface should be
colorless (i.e.\ color confined).  This is actually one of the known
properties fulfilled by a certain color-superconducting phase, that
is, the color-flavor locked (CFL)
phase~\cite{Schafer:1998ef,Alford:1999pa,Fukushima:2004bj,%
Hatsuda:2006ps}.  To form a colorless object in the CFL phase,
however, we need to treat a meson composed of four quarks, which is
technically complicated.  Instead of real QCD, here, we shall make use
of the two-color system;  this exhibits superfluidity at high density
which is reminiscent of color superconductivity in QCD, and besides,
we need not treat four quarks, since diquarks (that is, baryons in the
two-color world) can make a color singlet.  We will demonstrate that
the realization of quarkyonic matter is not so exclusive to disfavor
superfluidity and diquark condensation.  The goal of our discussions
in this part is to make convincing the phase diagram as drawn in
Fig.~\ref{fig:phase}.

\vspace{2mm}
Let us now specify the model that we will use.  Physics of Cooper
pairing of quarks (dense fermionic matter in general) is well
described by NJL-type effective models.  However, the NJL-type models
suffer from a serious drawback: the lack of confinement, which may
bring about artificial model predictions.  In order to capture the
essential features of confinement physics and yet maintain the
technical simplicity of the NJL model, it was augmented with the
Polyakov loop, equipped with a phenomenological potential designed to
reproduce selected lattice data at finite $T$ and zero
$\mu_B$~\cite{Fukushima:2003fw,Ratti:2005jh,Roessner:2006xn,Hansen:2006ee,%
Sasaki:2006ww,Rossner:2007ik,Abuki:2008ht,Fukushima:2008wg}.  (See
also, Ref.~\cite{Schaefer:2007pw} for example, for a related approach of
the Polyakov loop extended quark-meson model.)  This extended model is
now called the PNJL model.  In spite of its simplicity, the PNJL model
has shown remarkable agreement with thermodynamic quantities measured
in the lattice QCD simulations.  So far, among available finite-$T$
and finite-$\mu_B$ model options at least, the PNJL model is one of
the best tools to unveil the QCD phase structure.  In this paper, in a
sense, we \emph{downgrade} the PNJL model into the two-color setting,
hoping that our model predictions would guide the future two-color
simulations.

Figure~\ref{fig:phase} is actually the phase diagram of two-color two-flavor QCD
predicted by means of the PNJL model.  We note that the phase diagram is divided
into three regimes:  One at high $T$ consists
of deconfined and chiral symmetric particles, which is to be
identified as a quark-gluon plasma.  The other one at low $T$ and low
$\mu_B$ is of course the hadronic phase.  The last one at low $T$ and
high $\mu_B$ is commonly referred to as a superfluid state.  We will
later discuss that this state can be regarded as quarkyonic matter --
so to speak, quarkyonic superfluid of two-color quark matter.  Keeping
in mind this phase structure, we will look at the pole and screening
masses of mesons to extract the information on $\gr{U(1)_A}$
symmetry.  The meson spectrum in the PNJL model was first analyzed in
Ref.~\cite{Hansen:2006ee}.  Along the same line, we will perform the
calculations and examine the dependence of meson spectrum on the
$\gr{U(1)_A}$-breaking interaction strength.

The paper is organized as follows.  In Sec.~\ref{Sec:model} we
introduce the model Lagrangian and derive some basic analytic
formulas.  Before going into the numerical study, we describe in
Sec.~\ref{Sec:parfix} in detail the way we fix the parameters of our
model.  Most of the results have been obtained numerically and are
presented in Sec.~\ref{Sec:results}.  Finally, in
Sec.~\ref{Sec:conclusions} we summarize and conclude.


\section{Model Setup}
\label{Sec:model}

Two-color QCD with $N_f$ massless quark flavors has a global $\gr{U}(2N_f)$
flavor invariance at the classical level owing to the Pauli--G\"ursey
symmetry~\cite{Smilga:1995tb}.
The axial anomaly explicitly breaks $\gr{U}(2N_f)$ to $\gr{SU}(2N_f)$.
In the vacuum, the $\gr{SU}(2N_f)$ symmetry is spontaneously broken by the
standard chiral condensate down to its $\gr{Sp}(2N_f)$ subgroup.
In the $N_f=2$ case, which is the subject of the present paper, the
symmetry-breaking pattern can be equivalently cast as
$\gr{SO(6)\to SO(5)}$~\cite{Smilga:1995tb,Rapp:1998zu}.  The spectrum
of NG modes therefore consists of a single 5-plet, including three
pions and two diquarks (a diquark and an antidiquark -- a baryonic and
an antibaryonic pion).  Within the NJL model, this degeneracy is
reflected by the equality of couplings in the meson and diquark
channels~\cite{Ratti:2004ra}.

Let us begin the NJL analysis with the $\gr{U(1)_A}$-invariant interaction
Lagrangian,
\begin{multline}
 \mathcal L_1 = (1-\alpha)G \Bigl[ (\adj\psi\psi)^2
  + (\adj\psi i\gamma_5\vec\tau\psi)^2
  + (\adj\psi i\gamma_5\psi)^2 + (\adj\psi\vec\tau\psi)^2 \\
  + |\adj{\cc\psi}\gamma_5\sigma_2\tau_2\psi|^2
  + |\adj{\cc\psi}\sigma_2\tau_2\psi|^2
  \Bigr] \,,
\end{multline}
where $\vec\sigma$ and $\vec\tau$ denote Pauli matrices in color and
flavor spaces, and $\cc\psi$ the charge conjugation of the Dirac spinor
$\cc\psi=C\adj\psi^T$ with $C=i\gamma^2\gamma^0$.  The interaction $\mathcal
L_1$ is minimal in the sense that it only involves the scalar and pseudoscalar
channels with isospin zero and unity. While the above $\mathcal{L}_1$ is
invariant under a $\gr{U(1)_A}$ rotation, we further need an interaction which
breaks the $\gr{U(1)_A}$ symmetry. To that end we consider the analogous
interaction as follows:
\begin{multline}
 \mathcal L_2 = \alpha\,G \Bigl[ (\adj\psi\psi)^2
  + (\adj\psi i\gamma_5\vec\tau\psi)^2
  - (\adj\psi i\gamma_5\psi)^2 - (\adj\psi\vec\tau\psi)^2 \\
  + |\adj{\cc\psi}\gamma_5\sigma_2\tau_2\psi|^2
  - |\adj{\cc\psi}\sigma_2\tau_2\psi|^2
  \Bigr] \,.
\end{multline}
The general interaction Lagrangian is thus a sum of these two;
$\mathcal{L}_{\text{int}}=\mathcal{L}_1+\mathcal{L}_2$.  At $\alpha=0$
only $\mathcal{L}_1$ remains and the interaction preserves
$\gr{U(1)_A}$, whereas at $\alpha=1$ the remaining piece
$\mathcal{L}_2$ breaks $\gr{U(1)_A}$ maximally, being equivalent to
the two-flavor instanton-induced interaction~\cite{Rapp:1998zu}.
Indeed $\alpha$ is a usually used $\gr{U(1)_A}$-violating
parameter~\cite{Buballa:2003qv} but we will also use
$\zeta\equiv 1-2\alpha$ for notation simplicity.
In previous works, $\zeta=0$ ($\alpha=1/2$) was used to discuss the phase
structure~\cite{Ratti:2004ra,Sun:2007fc}.

Performing the Hubbard--Stratonovich transformation in all six
channels, we arrive at the total Lagrangian as
\begin{multline}
\mathcal L=\adj\psi \bigl(i\gamma^\mu
D_\mu-m_0+\gamma_0\mu-\sigma-i\gamma_5\vec\tau\cdot\vec\pi
-i\zeta\gamma_5\eta-\zeta\vec\tau\cdot\vec a \bigr)\psi\\
+\frac12 \Bigl(\Delta^*\adj{\cc\psi}i\gamma_5\sigma_2\tau_2\psi
 +\text{h.c.}\Bigr)
+\frac\zeta2 \Bigl(\Delta_5^*\adj{\cc\psi}i\sigma_2\tau_2\psi
 +\text{h.c.}\Bigr)\\
-\frac1{4G}\bigl(\sigma^2+\vec\pi^2+\zeta\eta^2+\zeta\vec a^2
+|\Delta|^2+\zeta|\Delta_5|^2 \bigr) \,.
\label{L_total}
\end{multline}
The covariant derivative $D_\mu$ involves coupling of the quarks to
the background gauge field $A_4$ which translates to the Polyakov loop
in the end.  In addition, we note that $m_0$ and $\mu$ denote the
current quark mass and the quark chemical potential.  Later we will
introduce $\mu_B$ to denote the baryon chemical potential;
$\mu_B=2\mu$ where 2 comes from the number of colors.  The collective
fields $\sigma,\vec\pi,\eta,\vec a,\Delta,\Delta_5$ represent in order
the mesons in the scalar-isoscalar, pseudoscalar-isovector,
pseudoscalar-isoscalar, and scalar-isovector channels, and the scalar
and pseudoscalar diquarks.  (We hereafter omit the subscript ``0'' out
of $\eta_0$ and $\vec{a}_0$ for simplicity.)

In the absence of isospin chemical potential the isovector modes do not
develop a vacuum expectation value. Moreover, the Vafa--Witten
theorem~\cite{Vafa:1984xg} gua\-ran\-tees that the chiral condensate in the
vacuum has positive parity. In our model approach, this requires
$\alpha\geq0$, that is, $\zeta\leq1$. (At the same time, $\zeta\geq0$, i.e.,
$\alpha\leq1/2$ is needed for our mean-field treatment using
the Hubbard--Stratonovich transformation \cite{Boer:2008ct}.) Consequently,
with the exception of the $\gr{U(1)_A}$-conserving limit $\alpha=0$, the scalar
chiral and diquark condensates will always be preferred to the pseudoscalar
ones. We will therefore take into account only the $\sigma$ and $\Delta$
condensates. The mean-field thermodynamics of the system is then independent of
the parameter $\alpha$, which will only affect the propagation of collective
modes, to be discussed in Sec.~\ref{Sec:collective}.


\subsection{Thermodynamics}

In the PNJL model, one introduces a constant temporal gauge field
which couples to the quarks via the covariant derivative.  In the
Polyakov gauge this gauge field is diagonal in the color space, and
for the color $\gr{SU(2)}$ group it has a form,
$A_4=iA_0=\sigma_3\theta$, where $\theta$ is a real ``phase''.  The
order parameter for deconfinement is then the traced Polyakov loop
given by
\begin{equation}
 \Phi=\frac1{N_c}\,\mathrm{Tr}\,e^{i\beta A_4}=\cos(\beta\theta) \,,
\end{equation}
where $\beta$ is the inverse temperature.  In the mean-field
approximation, the thermodynamic potential is given by a sum of the
gauge and quark parts,
\begin{equation}
 \Omega=\Omega_{\text{gauge}}+\Omega_{\text{quark}} \,.
\end{equation}
In the following, we will refer to the two quark colors for simplicity as
the ``red'' and ``green''. Combining the red quark and the green antiquark
into the Nambu--Gor'kov spinor, $\Psi=(\psi_r,\tau_2\cc\psi_g)^T$, the
background gauge field becomes proportional to the unit matrix in this
doubled space and the quark thermodynamic potential can be expressed
as
\begin{multline}
 \Omega_{\text{quark}} = \frac{\sigma^2+\Delta^2}{4G} \\
 -T\sum_n\int \frac{d^3\vek k}{(2\pi)^3}\,\mathrm{Tr}\,
 \log(i\omega_n-i\theta-\mathcal H_{\vek k}) \,,
\end{multline}
where the Nambu--Gor'kov Hamiltonian reads
\begin{equation}
 \mathcal H_{\vek k}=\begin{pmatrix}
 \vek\alpha\cdot\vek k+\gamma_0M-\mu & -\gamma_0\gamma_5\Delta\\
 \gamma_0\gamma_5\Delta^* & \vek\alpha\cdot\vek k+\gamma_0M+\mu
 \end{pmatrix} \,.
\end{equation}
Here $M=m_0+\sigma$ is the constituent quark mass and the $\sigma$ and
$\Delta$ now stand for the condensates. The four eigenvalues of the
Hamiltonian are easily found as $+E^\pm_{\vek k}$ and
$-E^\pm_{\vek k}$ corresponding to the gapped quasiparticle dispersion
relations, where
\begin{equation}
 \begin{split}
 & E^\pm_{\vek k} = \sqrt{(\xi^\pm_{\vek k})^2+\Delta^2} \,,\\
 & \xi^\pm_{\vek k} = \epsilon_{\vek k} \pm \mu \,,\qquad
   \epsilon_{\vek k} = \sqrt{\vek k^2 + M^2} \,.
 \end{split}
\end{equation}
The total thermodynamic potential thus becomes
\begin{equation}
 \begin{split}
 & \Omega = -bT\bigl[24\Phi^2 e^{-\beta a}+\log(1-\Phi^2)\bigr]
  +\frac{\sigma^2+\Delta^2}{4G} \\
 &- 4\sum_{i=\pm}\int\!\frac{d^3\vek k}{(2\pi)^3} \biggl[
  E^i_{\vek k} \! + \! T\log\left( 1 \! + \! 2\Phi
  e^{-\beta E^i_{\vek k}} \! + \! e^{-2\beta E^i_{\vek k}}
  \right)\biggr] \,.
 \end{split}
\label{TDpot}
\end{equation}
The first term is the gauge part $\Omega_{\text{gauge}}$ having two
model parameters $a$ and $b$.  We assume the simple form motivated by
lattice strong-coupling expansion~\cite{Fukushima:2008wg}.  It differs
from the three-color expression by a simpler logarithmic term due to
the $\gr{SU(2)}$ Haar measure, and by the rescaled prefactor of the
exponential term, which is in general proportional to $N_c^2$. Note that, as
usual in the PNJL model literature, we simulate the effects of gauge dynamics
by a constant background temporal gauge field. We then adopt a phenomenological
ansatz for the gauge contribution to the mean-field thermodynamic potential,
chosen to reproduce selected features of the pure gauge theory. Therefore, the
parameters $a,b$ only enter the thermodynamic potential \eqref{TDpot} since
there are no dynamical gauge degrees of freedom in our model Lagrangian
\eqref{L_total}.

It is interesting to recall that in the three-color case the thermodynamic
potential in general cannot be written in terms of the Polyakov loop variable
$\Phi$ (and the conjugate $\adj\Phi$) only, and one has to use two phases
analogous to our $\theta$ to parameterize it. On the contrary, in the two-color
PNJL model the thermodynamic potential depends just on $\Phi$ even in the
presence of a diquark condensate. This is because diquarks are colorless. This
considerably simplifies the discussion and also avoids technical ambiguities
stemming from generally complex effective actions involving the diquark
condensate~\cite{Roessner:2006xn,Rossner:2007ik,Abuki:2008ht,Abuki:2009dt}.
The values of the condensates in thermodynamic equilibrium are
determined by minimizing the thermodynamic potential with respect to
the variables $\sigma$, $\Delta$, and $\Phi$.

In the quark sector, the effect of the Polyakov loop as compared to
the simple NJL model is to make the replacement
$E+2T\log(1+e^{-\beta E})\to
E+T\log(1+2\Phi e^{-\beta E}+e^{-2\beta E})$ in the quasiparticle
contribution to the thermodynamic potential.  Similarly, in the gap
equations as well as collective mode propagators, one generalizes
$1-2f(E)=\tanh(\beta E/2)$, where $f(E)=1/(e^{\beta E}+1)$ is the
Fermi--Dirac distribution, to
\begin{align}
 \varphi(E) &\equiv \frac{\sinh(\beta E)}{\cosh(\beta E)+\Phi} \notag\\
 &= \left(1+\frac{1-\Phi}{\cosh(\beta E)+\Phi}\right)
 \bigl[ 1-2f(E) \bigr]
\label{eq:mod_f}\\
 &= \left(1-\frac{\Phi}{\cosh(\beta E)+\Phi}\right)
 \bigl[ 1-2f(2E) \bigr].
\label{eq:mod_f2}
\end{align}
The latter two forms (\ref{eq:mod_f}) and (\ref{eq:mod_f2}) of the
function $\varphi(x)$ illustrate that for $\Phi\to1$ thermal
excitations are dominated by quark modes with baryon number $1/2$
(i.e.\ deconfinement), while for $\Phi\to0$ by baryon modes with
baryon number $1$ (i.e.\ confinement).  With these replacements, one
can readily generalize the results of the pure NJL model to the
Polyakov loop-extended one, as observed in Ref.~\cite{Hansen:2006ee}
for the three-color case.

At zero temperature the Polyakov loop expectation value is zero for
all values of the chemical potential. This led to the suggestion that
the PNJL model can naturally describe quarkyonic
matter~\cite{McLerran:2007qj,Hidaka:2008yy} at high chemical
potential, that is, a phase where chiral symmetry is restored but
confinement
persists~\cite{Fukushima:2008wg,Schaefer:2007pw,Abuki:2008nm}.  A
simple glance at Eq.~\eqref{TDpot} shows that when $\Phi=0$, thermal
excitations are indeed governed by the term
$e^{-2\beta E^e_{\vek k}}$, i.e., they correspond to colorless
baryons. Let us take a closer look at how $\Phi\simeq0$ arises as $T\to0$.

We consider the gap equation for $\Phi$ following from
Eq.~\eqref{TDpot}, that is,
\begin{equation}
 b\Phi\left(\frac1{1-\Phi^2}-24e^{-\beta a}\right)
 = \sum_{i=\pm}\int\!\frac{d^3\vek k}{(2\pi)^3}
 \frac{2}{\Phi+\cosh(\beta E^i_{\vek k})} \,.
\end{equation}
At high enough $\mu$ and low $T$ the system is in the
Bardeen--Cooper--Schrieffer (BCS) regime where
Cooper pairing of quarks occurs close to the Fermi
sea~\cite{Sun:2007fc}.  The right-hand side of this equation is then
dominated by the particle ($i=-$) part.  We can further simplify the
calculation by using the high-density approximation, in which we
expand the dispersion relation around the Fermi surface,
$E^-=\sqrt{\xi^2+\Delta^2}\approx\Delta+\xi^2/(2\Delta)$, and replace
the measure $d^3\vek k/(2\pi)^3$ by $\mathcal N\,d\xi$, where
$\mathcal N=\mu k_{\text F}/(2\pi^2)$ is the density of states at the
Fermi surface.  The integral thus becomes Gaussian near $T\simeq0$ and
we arrive at the asymptotic result as
\begin{equation}
 \Phi_{\text{BCS}} \approx \frac{4\mathcal N}{b} \,
 e^{-\beta\Delta} \, \sqrt{2\pi\Delta T} \,.
\end{equation}
The Polyakov loop itself is hence suppressed exponentially at low
temperature.


\subsection{Collective modes}
\label{Sec:collective}

The propagators of the collective modes are easily obtained by a
second variation of the effective action that follows from
Eq.~\eqref{L_total} after integrating the quarks fields out.  Denoting
a set of collective fields symbolically as $\chi_i$, the inverse
propagator at imaginary (bosonic Matsubara) frequency $i\omega'_m$ is given by
\begin{multline}
 D_{ij}^{-1}(i\omega'_m,\vek p) = C_i\delta_{ij}+T\sum_n\int\!
  \frac{d^3\vek k}{(2\pi)^3}\\
 \times\mathrm{Tr}\left[\frac{\partial\mathcal H}{\partial\chi_i}
  \frac{1}{i(\tilde\omega_n \!+\! \omega'_m) \!-\!
  \mathcal H_{\vek k+\frac{\vek p}{2}}}
  \frac{\partial\mathcal H}{\partial\chi_j}
  \frac{1}{i\tilde\omega_n-\mathcal H_{\vek k \!- \! \frac{\vek p}{2}}}
  \right] ,
\end{multline}
where $\tilde\omega_n=\omega_n-\theta$ and $\omega_n$ stands for the fermionic
Matsubara frequencies. The constant $C_i$ is equal to $1/(2G)$ for
$\sigma,\vec\pi$, to $1/(4G)$ for $\Delta,\Delta^*$,
to $\zeta/(2G)$ for $\eta,\vec a$, and to $\zeta/(4G)$ for
$\Delta_5,\Delta_5^*$.  The trace is taken in Dirac, flavor, as well
as Nambu--Gor'kov space.  When calculating the partial derivatives of
the Hamiltonian, that determine the Yukawa couplings of quarks to the
bosonic modes, we demand that all fields have to be kept as in
Eq.~\eqref{L_total}.

In the diquark condensation phase, the baryon number is spontaneously
broken.  As a result some of the modes mix and we have to find their
dispersion relations by diagonalizing a matrix propagator.  This
applies to the scalar-isoscalar modes, $\sigma,\Delta,\Delta^*$, as
well as pseudoscalar-isoscalar modes, $\eta,\Delta_5,\Delta_5^*$.  The
only modes that do not mix are $\vec\pi$ and $\vec a$ thanks to
conservation of isospin and parity.  Their inverse propagators are
trivial in isospin space, namely, $D^{-1}_{ij}=\delta_{ij}D^{-1}_i$.
After analytic continuation to real frequencies they acquire the
following forms:
\begin{widetext}
\begin{equation}
\begin{split}
D^{-1}_\pi(\omega,\vek p)=&\frac{1}{2G}-\sum_{i,j=\pm}\sum_{k,l=\pm}
\int\!\frac{d^3\vek q}{(2\pi)^3}\,
 \frac1{2E^i_{\vek q+\frac{\vek p}2} E^{-j}_{\vek q-\frac{\vek p}2}}
 \left(1+ij\frac{\epsilon_{\vek q}^2-\frac{\vek p^2}{4}}
 {\epsilon_{\vek q+\frac{\vek p}{2}} \,
  \epsilon_{\vek q-\frac{\vek p}{2}}}\right)\\
&\times\frac{(E^i_{\vek q+\frac{\vek p}2}
 +ik\,\xi^i_{\vek q+\frac{\vek p}2})
 (E^{-j}_{\vek q-\frac{\vek p}2}
 +jl\,\xi^{-j}_{\vek q-\frac{\vek p}2})
 +kl\Delta^2}  {\omega+k E^i_{\vek q+\frac{\vek p}2}
 +l E^{-j}_{\vek q-\frac{\vek p}2}}
 \Bigl[\varphi(k E^i_{\vek q+\frac{\vek p}2})
 +\varphi(l E^{-j}_{\vek q-\frac{\vek p}2})\Bigr],\\
D^{-1}_a(\omega,\vek p)=&\frac{\zeta}{2G}-\zeta^2\sum_{i,j=\pm}
\sum_{k,l=\pm} \int\!\frac{d^3\vek q}{(2\pi)^3}
 \frac1{2E^i_{\vek q+\frac{\vek p}2} E^{-j}_{\vek q-\frac{\vek p}2}}
 \left(1-ij\frac{M^2-\vek q^2+\frac{\vek p^2}{4}}
 {\epsilon_{\vek q+ \frac{\vek p}{2}} \,
  \epsilon_{\vek q-\frac{\vek p}{2}}}\right)\\
&\times\frac{(E^i_{\vek q+\frac{\vek p}2}+ik\,
 \xi^i_{\vek q+\frac{\vek p}2})(E^{-j}_{\vek q-\frac{\vek p}2}
 +jl\,\xi^{-j}_{\vek q-\frac{\vek p}2})-kl\Delta^2}
 {\omega+k E^i_{\vek q+\frac{\vek p}2}
 + l E^{-j}_{\vek q-\frac{\vek p}2}}
 \Bigl[\varphi(k E^i_{\vek q+\frac{\vek p}2})
 +\varphi(l E^{-j}_{\vek q-\frac{\vek p}2})\Bigr] \,.
\end{split}
\label{medium_props}
\end{equation}
Here, we have changed the momentum notation from $\vek k$ to $\vek q$
to reserve $k$ to take a summation over $\pm$.  It should be noted
that in the limit $|\vek p|\to0$, only the $i=j$ terms survive and the
expressions somewhat simplify.  The first of the formulas can be used
together with the gap equation for $\Delta$ to show that in the
diquark condensation phase the pion (pole) mass is exactly equal to
$2\mu=\mu_B$ at zero temperature.  The proof is straightforward, hence we omit the
details.

In the normal phase (where $\Delta=0$) all propagators can be
evaluated easily.  We provide here the list of expressions that we
later in Sec.~\ref{Sec:num_collective} use to calculate the masses
numerically.  To recall that these are the propagators in normal
matter, let us write a superscript $(n)$:
\begin{equation}
\begin{split}
D^{(n)-1}_\sigma(\omega,\vek p) &= \frac{1}{2G}-2\sum_{i,j=\pm}
 \int\!\frac{d^3\vek k}{(2\pi)^3}
  \left(1-ij\frac{M^2-\vek k^2+\frac{\vek p^2}{4}}
  {\epsilon_{\vek k+\frac{\vek p}{2}} \,
   \epsilon_{\vek k-\frac{\vek p}{2}}}\right)
  \frac{\varphi(i\,\xi^i_{\vek k+\frac{\vek p}{2}})+
        \varphi(j\,\xi^{-j}_{\vek k-\frac{\vek p}{2}})}
   {\omega+i\,\epsilon_{\vek k+\frac{\vek p}{2}}+
    j\,\epsilon_{\vek k-\frac{\vek p}{2}}} \,,\\
D^{(n)-1}_\pi(\omega,\vek p) &= \frac{1}{2G}-2\sum_{i,j=\pm}
 \int\!\frac{d^3\vek k}{(2\pi)^3}
  \left(1+ij\frac{\epsilon_{\vek k}^2-\frac{\vek p^2}{4}}
  {\epsilon_{\vek k+\frac{\vek p}{2}} \,
   \epsilon_{\vek k-\frac{\vek p}{2}}}\right)
  \frac{\varphi(i\,\xi^i_{\vek k+\frac{\vek p}{2}})+
        \varphi(j\,\xi^{-j}_{\vek k-\frac{\vek p}{2}})}
   {\omega+i\,\epsilon_{\vek k+\frac{\vek p}{2}}+
    j\,\epsilon_{\vek k-\frac{\vek p}{2}}} \,,\\
D^{(n)-1}_\eta(\omega,\vek p) &= \frac{\zeta}{2G}-2\zeta^2\sum_{i,j=\pm}
 \int\!\frac{d^3\vek k}{(2\pi)^3}
  \left(1+ij\frac{\epsilon_{\vek k}^2-\frac{\vek p^2}{4}}
  {\epsilon_{\vek k+\frac{\vek p}{2}} \,
   \epsilon_{\vek k-\frac{\vek p}{2}}}\right)
  \frac{\varphi(i\, \xi^i_{\vek k+\frac{\vek p}{2}})+
        \varphi(j\, \xi^{-j}_{\vek k-\frac{\vek p}{2}})}
   {\omega+i\,\epsilon_{\vek k+\frac{\vek p}{2}}+
    j\,\epsilon_{\vek k-\frac{\vek p}{2}}} \,,\\
D^{(n)-1}_a(\omega,\vek p) &= \frac{\zeta}{2G}-2\zeta^2\sum_{i,j=\pm}
 \int\!\frac{d^3\vek k}{(2\pi)^3}
  \left(1-ij\frac{M^2-\vek k^2+\frac{\vek p^2}{4}}
  {\epsilon_{\vek k+\frac{\vek p}{2}} \,
   \epsilon_{\vek k-\frac{\vek p}{2}}}\right)
  \frac{\varphi(i\, \xi^i_{\vek k+\frac{\vek p}{2}})+
        \varphi(j\, \xi^{-j}_{\vek k-\frac{\vek p}{2}})}
   {\omega+i\,\epsilon_{\vek k+\frac{\vek p}{2}}+
    j\,\epsilon_{\vek k-\frac{\vek p}{2}}} \,,\\
D^{(n)-1}_\Delta(\omega,\vek p) &= \frac{1}{4G}-\sum_{i,j=\pm}
 \int\!\frac{d^3\vek k}{(2\pi)^3}
  \left(1+ij\frac{\epsilon_{\vek k}^2-\frac{\vek p^2}{4}}
  {\epsilon_{\vek k+\frac{\vek p}{2}} \,
   \epsilon_{\vek k-\frac{\vek p}{2}}}\right)
  \frac{\varphi(i\, \xi^i_{\vek k+\frac{\vek p}{2}})+
        \varphi(j\, \xi^j_{\vek k-\frac{\vek p}{2}})}
  {\omega+2\mu+i\,\epsilon_{\vek k+\frac{\vek p}{2}}+
   j\,\epsilon_{\vek k-\frac{\vek p}{2}}} \,,\\
D^{(n)-1}_{\Delta_5}(\omega,\vek p) &= \frac{\zeta}{4G}-\zeta^2\sum_{i,j=\pm}
 \int\!\frac{d^3\vek k}{(2\pi)^3}
  \left(1-ij\frac{M^2-\vek k^2+\frac{\vek p^2}{4}}
  {\epsilon_{\vek k+\frac{\vek p}{2}} \,
   \epsilon_{\vek k-\frac{\vek p}{2}}}\right)
  \frac{\varphi(i\, \xi^i_{\vek k+\frac{\vek p}{2}})+
        \varphi(j\, \xi^j_{\vek k-\frac{\vek p}{2}})}
  {\omega+2\mu+i\,\epsilon_{\vek k+\frac{\vek p}{2}}+
   j\,\epsilon_{\vek k-\frac{\vek p}{2}}} \,.
\end{split}
\label{full_props}
\end{equation}
\end{widetext}
Propagators of the antidiquarks, $\Delta^*$ and $\Delta_5^*$, are
obtained from the diquark ones by charge conjugation, i.e.\ changing
the sign of the chemical potential.

\begin{table*}
\caption{First two lines: physical quantities used as an input. Last
  two lines: fitted parameters of the model.}
\label{Tab:parameters}
\includegraphics[scale=1]{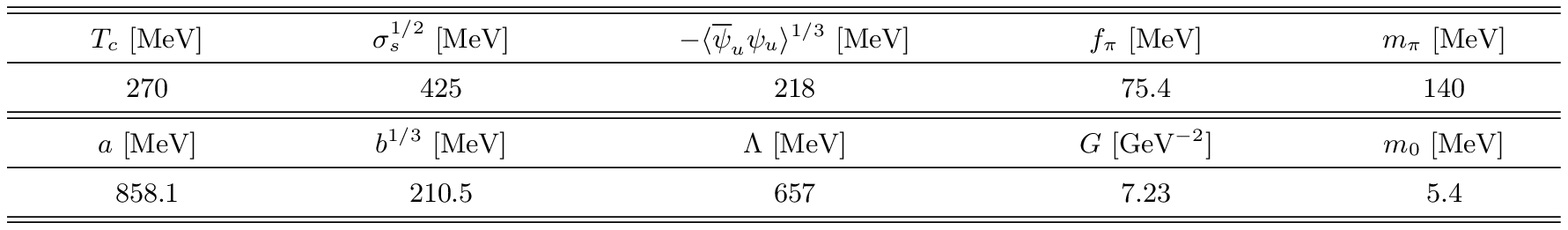}
\end{table*}


\section{Numerical Results}
\label{Sec:results}

Armed with these analytic expressions we are now ready to proceed in
the numerical calculations.  First we will explain our choice of the
model parameters.  Next we will investigate the phase diagram, in which
we particularly look into the nature of the superfluid phase.  Finally
we will come to the discussion of the meson spectrum dependence on the
$\gr{U(1)_A}$-breaking parameter $\alpha$.


\subsection{Parameter fixing}
\label{Sec:parfix}

Our model has five parameters whose values have to be chosen
appropriately: $a$ and $b$ in the Polyakov loop sector, and $G$,
$m_0$, and the sharp three-momentum cutoff $\Lambda$ in the NJL
sector. (The parameter $\alpha$ or $\zeta$ will be treated as an
undetermined free parameter.)

Let us first concentrate on the latter.  In the three-color NJL model one
normally fixes the values of $G$, $m_0$, and $\Lambda$ from the
physical pion mass $m_\pi$, decay constant $f_\pi$, which are both
measured experimentally, and the chiral condensate, which is
calculated on the lattice or estimated from the QCD sum
rules~\cite{Buballa:2003qv}.  The authors of Ref.~\cite{Ratti:2004ra}
chose to set up the two-color NJL model using roughly the same input
values, while in Ref.~\cite{Sun:2007fc} the pion mass, decay constant
and the (three-color) constituent quark mass were used, resulting in a
rather different parameter set.  We shall argue here that both these
fits may be actually overdetermined and one should be very careful
especially when comparing the model outputs with lattice data.

The chiral limit QCD, as a Yang--Mills theory coupled to massless
quarks in the color fundamental representation, has a single mass
scale, namely $\Lambda_{\text{QCD}}$.  All physical quantities should
be expressed as an appropriate power of $\Lambda_{\text{QCD}}$ times a
dimensionless number.  In the NJL model one can therefore choose
freely a single input quantity, basically just to set the energy unit.
Dimensionless combinations of observables are a pure prediction of
QCD, and shall thus not be tuned to arbitrary values.  In particular,
either the pion decay constant, the chiral condensate, or the
constituent quark mass may serve for this purpose, but not two of them
independently.  With the fact in mind that the current quark mass is a
free parameter in the lattice simulation, we note that the pion mass
can in principle acquire any value and represents a new physical scale
in addition to $\Lambda_{\text{QCD}}$.

\begin{figure}
\includegraphics[scale=1]{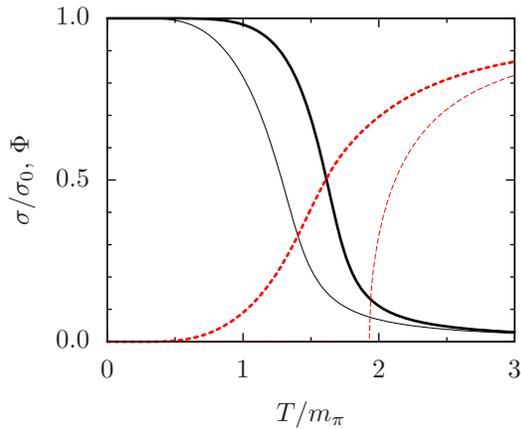}
\caption{Chiral and deconfinement crossovers at $\mu_B=0$.  The thick
  lines show the PNJL result (black solid: chiral condensate in units
  of the vacuum value $\sigma_0$; red dashed: Polyakov loop
  $\Phi$), while the thin lines indicate the pure chiral (NJL) and
  pure gauge transitions. As expected, the deconfinement transition in
  the absence of quarks is of second order.}
\label{Fig:mu0xovers}
\end{figure}

Unfortunately, we are not aware of any lattice data which would
provide us with the input needed to fix the NJL model parameters
unambiguously.  In order to make at least an educated guess, we use an
argument based on the $N_c$ scaling of physical quantities.  Using the
fact that $f_\pi$ is proportional to $\sqrt{N_c}$ and the chiral
condensate to $N_c$, we rescale the three-color values by factors
$\sqrt{2/3}$ and $2/3$, respectively.  Regarding the Polyakov loop
sector parameters, the constant $a$ is related to the critical
temperature $T_c$ for deconfinement in the pure gauge theory by
$a=T_c\log24$.  Since $T_c$ in the first approximation does not depend
on $N_c$~\cite{McLerran:2007qj}, we use the value
$T_c=270\;\text{MeV}$.  The parameter $b$ can in principle be adjusted
in order to make the chiral and deconfinement crossovers happen at
about the same temperature~\cite{Fukushima:2008wg}.  Here we use the
estimate based on lattice strong-coupling expansion,
$b=(\sigma_s/a)^3$, where $\sigma_s=(425\text{ MeV})^2$ is the
physical string tension.  The input values as well as the fitted
parameter set are summarized in Table~\ref{Tab:parameters}.

In order to check that our parameter set is reasonable, we plot in
Fig.~\ref{Fig:mu0xovers} the expectation values of $\sigma$ and $\Phi$
as a function of $T$ at $\mu_B=0$.  The positions of the two
crossovers move very close to each other when the coupling between the
quark and Polyakov loop sectors is switched on.  We note that this
observation of simultaneous crossovers is less clear if we use the
unrescaled input parameters.

\begin{figure*}
\includegraphics[scale=1]{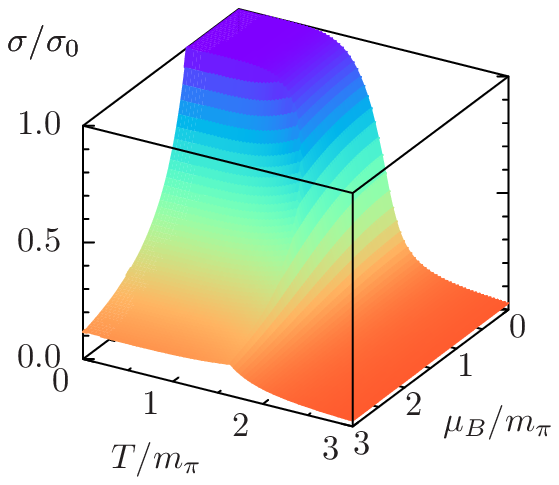}\hskip1em
\includegraphics[scale=1]{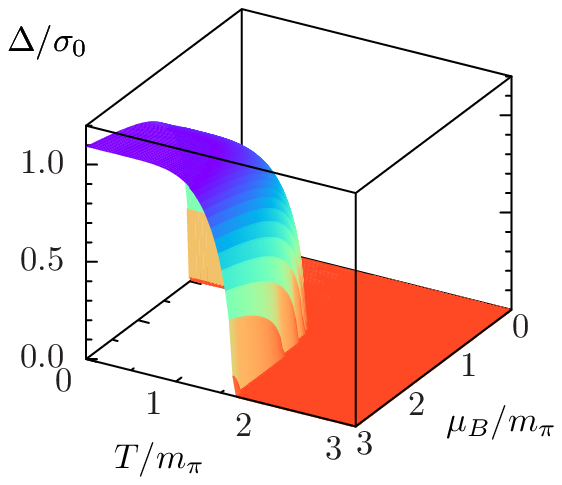}\hskip1em
\includegraphics[scale=1]{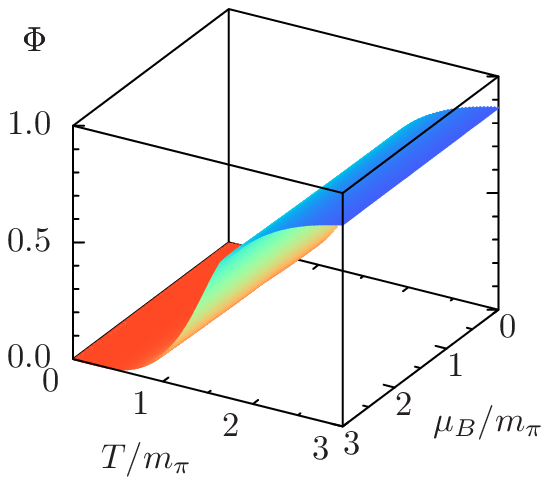}
\caption{Chiral condensate $\sigma$, diquark condensate $\Delta$,
  which are given as divided by the chiral condensate in the vacuum
  $\sigma_0$, and the Polyakov loop $\Phi$ as a function of $\mu_B$
  and $T$ in units of the pion mass $m_\pi=140\;\text{MeV}$. Note the
  orientation of the axes, chosen to obtain a better view of the surfaces!
}
\label{Fig:3Dcondensates}
\end{figure*}


\subsection{Phase diagram and diquark condensation}
\label{sec:phase}

In the vacuum the diquark is degenerate with the pions and its mass is therefore
$m_\pi$.  Thus, when $\mu_B(=2\mu)$ exceeds $m_\pi$, the diquark condenses and
one enters the BEC phase which forms a superfluid component.  At this moment the
constituent quark mass $M$ is still rather large.  However, as $\mu_B$ further
increases, $M$ drops.  Once $\mu>M$, a Fermi sea of quarks appears.  Here we
note that the baryon number density, $n_B$, acquires a contribution from
the diquark condensate and becomes nonzero as soon as
$\mu_B=m_\pi$ or $\mu=m_\pi/2$. Because of the binding energy, naturally,
$m_\pi/2$ is smaller than $M$, and thus the onset of $n_B\neq0$ emerges first
and then a quark Fermi sea shows up with increasing $\mu$.  In the presence of
the Fermi sea the diquark condensation is closer to the BCS pairing of quarks
sitting near the Fermi surface rather than to BEC of bound bosonic molecules.
There is no phase transition associated with this qualitative change of
behavior, so one speaks of a BCS--BEC crossover. Even though it is not
particularly sharp, it can be conveniently defined by the condition
$\mu=M$~\cite{Sun:2007fc}.

We will draw the phase diagram of our model in a ``conventional'' way.
In the phase diagram the deconfinement crossover is conveniently
defined by the condition $\Phi=0.5$.  The deconfinement temperature is
then almost independent of the chemical potential.  In fact, the value
of the Polyakov loop at all temperatures depends on $\mu_B$ very
weakly, as can be seen from Fig.~\ref{Fig:3Dcondensates}, where all
condensates are plotted as a function of $T/m_\pi$ and $\mu_B/m_\pi$
where $m_\pi$ is fixed to be $140\;\text{MeV}$.  This behavior of the
Polyakov loop can be traced back to the fact that the two-color
diquark is a color singlet, and therefore does not break center
symmetry to induce nonzero Polyakov loop directly.

Even at nonzero quark mass the diquark condensate exhibits a clear
second-order phase transition as shown in the plot for
$\Delta/\sigma_0$ (where $\sigma_0$ is the vacuum value of the chiral
condensate) in Fig.~\ref{Fig:3Dcondensates}.  Unlike the deconfinement
crossover, we can draw a well-defined phase boundary in the phase
diagram separating the normal and superfluid phases.  As we have
already mentioned above, we can confirm that nonzero $\Delta$
certainly appears at $\mu_B=m_\pi$ at $T=0$.

\begin{figure}
\includegraphics[scale=1]{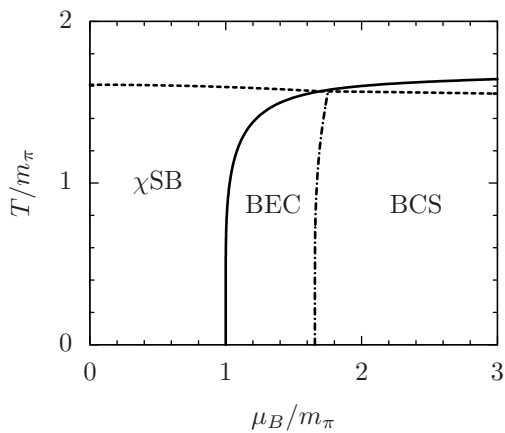}
\caption{Conventional presentation of the phase diagram of two-color
  QCD from the PNJL model in the $\mu_B-T$ plane.  We indicate the
  chiral-symmetry breaking ($\chi$SB), BEC, and BCS phases.  Solid
  line: critical onset of diquark condensation where $\Delta$ starts
  to be nonzero.  Dashed line: deconfinement crossover (defined by
  $\Phi=0.5$).  Dash-dotted line: BCS--BEC crossover (defined by
  $\mu=M$).
}
\label{Fig:phase_diagram}
\end{figure}

A compilation of these data leads to the phase diagram we present in
Fig.~\ref{Fig:phase_diagram}.  The solid line represents a
second-order phase transition between the normal and superfluid
phases.  In the superfluid region we have added a dash-dotted line
which indicates $\mu=M$ and can be interpreted as a BCS--BEC-type
crossover.  The dashed line is the deconfinement crossover defined by
$\Phi=0.5$.  Figure~\ref{Fig:phase_diagram} is a basis to understand
a more ``advocative'' way of presenting the phase diagram as shown in
Fig.~\ref{fig:phase}.

Now we are well prepared to discuss the physical meaning of each phase
labeled in Fig.~\ref{fig:phase}.  The red band which spreads almost
straight along the horizontal axis represents the deconfinement
crossover.  Because a crossover has a width and does not have a unique
definition, it should be much more reasonable to express the
transition region not by a line but a band.  The band width
in fact tells us how rapid or slow the crossover is.  We drew the
deconfinement band by the condition that $\Phi$ ranges from 0.4 to
0.6.  The blue band showing a sudden decrease around
$\mu_B/m_\pi\simeq 1.5$ is the chiral crossover defined similarly by
the condition that $\sigma/\sigma_0$ ranges from 0.4 to 0.6.  We see
that the two crossovers of deconfinement and chiral restoration take place
simultaneously at zero density, and this coincidence persists until
around $\mu_B/m_\pi\simeq 1.5$.  The pink dotted line represents the
superfluid onset, which is a well-defined phase transition.  Finally
the green band which shows behavior similar to the chiral
crossover is drawn by the baryon number density $n_B$.  We can compute
$n_B$ in the PNJL model and normalize it by the ``Stefan--Boltzmann''
value.  Since the PNJL model is a cutoff theory, even the free
(non-interacting) limit suffers from the cutoff artifact and deviates
from the standard formula, which is also the case in the lattice
simulation~\cite{Hegde:2008nx}.  Therefore we evaluate the baryon
number density in the Stefan--Boltzmann limit $(n_B)_{\rm SB}$ using
the PNJL model with $M=0$ and $\Phi=1$ imposed by hand.  In this way,
we indicate by the green band in Fig.~\ref{fig:phase} the region in
which $n_B/(n_B)_{\rm SB}$ ranges from 0.4 to 0.6.

Because the chiral phase transition controls the dynamical quark mass,
it is conceivable that the BCS--BEC crossover in Fig.~\ref{Fig:phase_diagram} is
associated with the chiral crossover. This is indeed the case;  the
dash-dotted line in Fig.~\ref{Fig:phase_diagram} is covered by the chiral
crossover band in Fig.~\ref{fig:phase}. Hence, as
labeled in Fig.~\ref{fig:phase}, the right-bottom region is characterized by
small $\Phi$ (confinement) and small $\sigma/\sigma_0$ (chiral symmetric).  This
is in fact in accord with the identification of quarkyonic matter in
Refs.~\cite{Fukushima:2008wg,Abuki:2008nm}.  In this case the blue band is
interpreted as the quarkyonic transition.  If we use $n_B/(n_B)_{\rm SB}$ to
define the quarkyonic transition according to
Refs.~\cite{McLerran:2007qj,McLerran:2008ua}, the green band, instead
of the blue band, plays the role of the quarkyonic boundary.

We would emphasize here that the former criterion makes more physical
sense at least in the present case of $N_c=2$.  Actually, our
consideration of the BCS--BEC crossover provides us with a clear view
point on this issue.  As we have already discussed, in the chemical
potential window $1\lesssim \mu_B/m_\pi \lesssim 1.5$, finite baryon
number density grows.  The carriers of the baryon number are, however,
not quarks but baryons (baryonic pions).  This is so because
$m_\pi<\mu_B<2M$ in this region.  Therefore, in the phase region we
called BEC in Fig.~\ref{Fig:phase_diagram}, the more appropriate
physical interpretation should be ``superfluid nuclear matter'' rather
than a quarkyonic state in which the pressure is mostly given by
Fermi-degenerated quarks.  This difference in the interpretation makes
a contrast to the large-$N_c$
arguments~\cite{McLerran:2007qj,McLerran:2008ua}, and it still remains
a question to which the real world with $N_c=3$ is closer, infinite
$N_c$ or $N_c=2$?  Answering this question goes beyond our current
scope.  If the quark-hadron continuity scenario driven by the CFL
state is realistic, we can say that the situation at $N_c=2$ is more
relevant.

Before closing this section we mention the previous studies.  In preceding works
there was some controversy regarding the order of the phase transition from the
diquark condensation phase to the normal phase. In~\cite{Ratti:2004ra} it
was concluded that there is a tricritical point at $\mu_B/m_\pi$ somewhere in
the range $2.2-2.4$, and for higher $\mu_B$ the transition becomes first order.
On the other hand, the authors of~\cite{Sun:2007fc} used the Thouless
criterion to calculate the critical temperature, which \emph{assumes} that the
transition is second order.  Our numerical results for the diquark condensate as
a function of $T/m_\pi$ and $\mu_B/m_\pi$ (see Fig.~\ref{Fig:3Dcondensates})
suggest that the transition is second order everywhere.  We have
further confirmed the statement that there is no first-order phase transition by
looking at the quartic Ginzburg--Landau coefficient in our numerical
calculation. Detailed computations and arguments are given in the Appendix.

\begin{figure}
\includegraphics[scale=1]{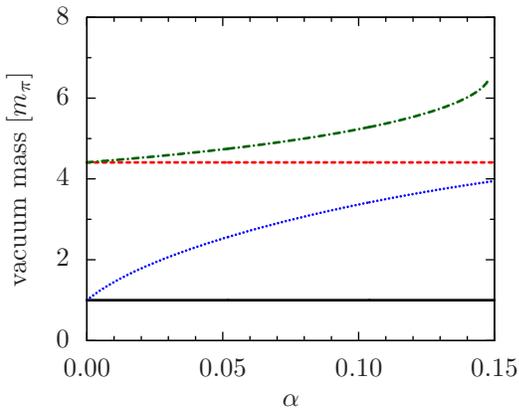}
\caption{Dependence of the vacuum meson masses on the
  $\gr{U(1)_A}$-breaking parameter $\alpha$.  The black solid is
  $m_\pi$, the red dashed $m_\sigma$, the blue dotted $m_\eta$, and
  the green dash-dotted $m_a$.  We see that $m_\pi=m_\eta$ and
  $m_\sigma=m_a$ in the $\gr{U(1)_A}$ symmetric ($\alpha=0$)
  case.}
\label{Fig:vacmass}
\end{figure}


\subsection{Collective mode spectrum}
\label{Sec:num_collective}

We are going to investigate the mass spectrum of collective modes as a function
of $T$ and $\mu_B$ for different values of the $\gr{U(1)_A}$-breaking parameter
$\alpha$.  A useful starting point therefore is the $\alpha$-dependence of the
masses in the vacuum. This is shown in Fig.~\ref{Fig:vacmass}.  It is worth
noting that we only display the meson spectrum; thanks to the unbroken
$\gr{SO(5)}$ symmetry at $\mu_B=0$, the scalar diquark $\Delta$ is degenerate
with $\pi$ mesons and the pseudoscalar diquark $\Delta_5$ is degenerate
with the $a_0$ mesons. We should perhaps emphasize that whenever we speak
of a (pole) mass, we have in mind the zero of the real part of the inverse
propagator~\cite{Hatsuda:1994pi}. We will therefore sometimes refer to it as
the real-part mass. This coincides with the position of the pole if the zero
appears below the threshold for decay into quark pairs.  Another prescription
for the mass would be to take the real part of the complex pole of the meson
propagator, yielding a somewhat different result, or to compute the spectral
function whose peak position and broadness indicate the physical mass and decay
width.

In the $\gr{U(1)_A}$-symmetric limit ($\alpha=0$), in our calculations,
the masses of $\eta_0$ and $a_0$ are equal to those of $\pi$ and $\sigma$,
respectively. The degeneracy of $\eta_0$ and $\pi$ is generally exact only when
the quark mass is strictly zero, so that they are both massless NG bosons. In
the (P)NJL model in the mean-field approximation, this exact degeneracy holds
regardless of finite quark masses, which is an artifact of the approximation.
In fact, $\eta_0$ and $\pi$ belong to different irreducible representations of
the unbroken $\gr{SO(5)}$;  $\eta_0$ is a singlet and $\pi$ (with $\Delta$) form
a quintet. The degeneracy is only approximate for small quark mass once
higher-order meson loops are taken into account.

Off the limit of $\alpha=0$, both $m_\eta$ and $m_a$ increase
steeply. The typical value of $\alpha$ in the three-color NJL model with two
light quark flavors, that one can obtain either by fitting the physical $\eta'$
mass or by a reduction of the three-flavor model, is in the range
$0.1-0.2$~\cite{Buballa:2003qv}. In order to be able to investigate the
convergence to the $\gr{U(1)_A}$-symmetric limit, we consider in the following
two particular values, $\alpha=0.05$ and $\alpha=0.1$.  That is, if we assume
that the vacuum value is $\alpha\simeq 0.1$, we consider two examples of no
reduction at all and $50\%$ reduction of $\gr{U(1)_A}$ effects.  It is quite
unlikely that $\alpha$ goes to zero in the $T$ and $\mu_B$ range of our
interest.

\begin{figure}
\includegraphics[scale=1]{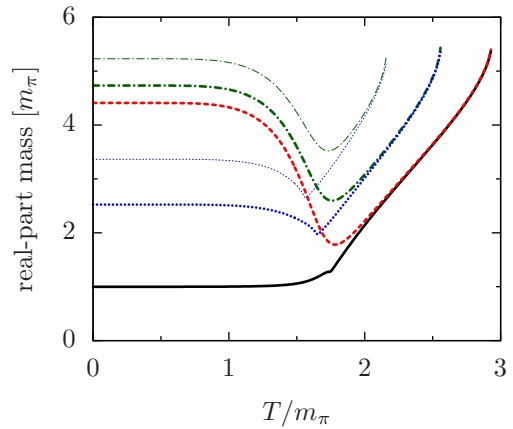}
\caption{Meson real-part masses at $\mu_B=0$ as a function of $T/m_\pi$.
  The notation for the lines is the same as in Fig.~\ref{Fig:vacmass}:
  The black solid is $m_\pi$, the red dashed $m_\sigma$, the blue
  dotted $m_\eta$, and the green dash-dotted $m_a$.  The thick
  lines correspond to $m_\eta$ and $m_a$ at $\alpha=0.05$, while the
  thin lines to $\alpha=0.1$.
}
\label{Fig:mu0mass}
\end{figure}
In Fig.~\ref{Fig:mu0mass} the meson real-part masses are plotted as a function
of $T/m_\pi$ while keeping $\mu_B=0$.  The reason why some curves exhibit
a cusp structure is that the corresponding modes cross the quark--antiquark (or
quark--quark in the case of diquarks) threshold as $T$ increases (and $M$
decreases accordingly).  Of course, such a decay into quarks is unphysical and
is just an artifact, following from the lack of confinement in the NJL model.
Even in the PNJL model these decay processes are not sufficiently suppressed
\cite{Hansen:2006ee}, since the coupling to the Polyakov loop only imitates
confinement in a statistical sense.

\begin{figure}
\includegraphics[scale=1]{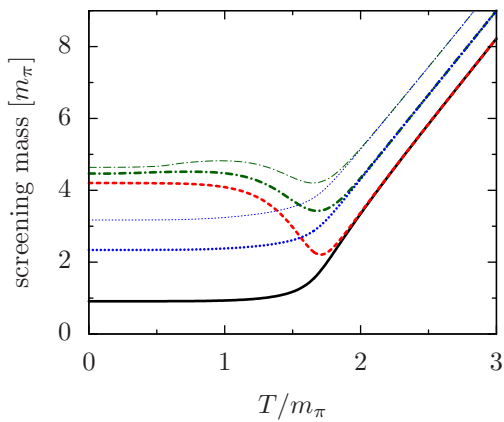}
\caption{Meson screening masses at $\mu_B=0$ as a function of
  $T/m_\pi$.  The notation for the lines is the same as in
  Fig.~\ref{Fig:vacmass}.  The black solid is $m_\pi$, the red dashed
  $m_\sigma$, the blue dotted $m_\eta$, and the green dash-dotted
  $m_a$.  The thick lines correspond to $m_\eta$ and $m_a$ at
  $\alpha=0.05$, while the thin lines to $\alpha=0.1$.}
\label{Fig:mu0screen}
\end{figure}

The real-part masses are difficult to measure on the lattice because the
number of lattice sites in the temporal direction is severely limited.
Instead, one can straightforwardly determine the damping of
correlators of the fermion-bilinear interpolating fields for the
mesons at large \emph{spatial} separations.  The exponential decay of
the correlations is related to the screening mass of the lightest mode
in the selected channel.  In the (P)NJL model, this can be found as
the pole of the static propagator, $D(\omega=0,\vek p)$, in the
complex-momentum plane.  The results are shown in
Fig.~\ref{Fig:mu0screen}.  At $T=0$ the real-part and screening masses
should coincide thanks to the Lorentz invariance.  In the model
calculation, however, this nice feature is slightly breached by the
three-momentum cutoff, but the difference of the two masses in the
vacuum turns out to be about a few percent at most.  Hence the
apparent cutoff artifacts are reasonably small.

The masses of the chiral partners become degenerate at high
temperature, signaling the restoration of chiral symmetry.  On the
other hand, at any (fixed) $\alpha\neq0$ the masses of the parity
partners, connected by a $\gr{U(1)_A}$ rotation, do not converge even
at the highest temperatures considered.  From the technical point of
view, this is a consequence of the simple structure of the inverse
propagators~\eqref{full_props}: when chiral symmetry is restored, the
self-energies of the parity partners become equal up to a simple
rescaling by $\zeta^2=(1-2\alpha)^2$.  Physically, in reality, one
should expect the coupling $\alpha$ to vary with $T$ since it
is induced by instantons whose density is exponentially suppressed at
high $T$~\cite{Gross:1980br}.  If we consider that the
$\gr{U(1)_A}$-breaking interaction strength is proportional to the
(pure) topological susceptibility, we can infer the $T$-dependence
from the lattice data in the pure gauge simulation.  Instead of doing
so, in this work, we pick up several values of $\alpha$.

A proper way to understand Fig.~\ref{Fig:mu0screen} is thus as
follows.  At $T=0$ naturally $\alpha$ is nonzero, and if precise
two-color simulation data is available for $m_\eta$ and $m_a$, in
principle, $\alpha$ can be fixed by the data.  We can perform the PNJL
model calculations using the determined $\alpha$ to go to the higher
temperature.  If we see a reduction of $m_\eta$ and $m_a$ toward
degenerated $m_\pi$ and $m_\sigma$, it is a signal for the effective
$\gr{U(1)_A}$ restoration.  We can deduce how far $\alpha$ decreases
by adjusting $\alpha$ to fit $m_\eta$ and $m_a$ at each temperature.
Hence, Fig.~\ref{Fig:mu0screen} is a demonstration for all these
possible investigations once the two-color simulation successfully
measures the screening masses in good precision.

\begin{figure}
\includegraphics[scale=1]{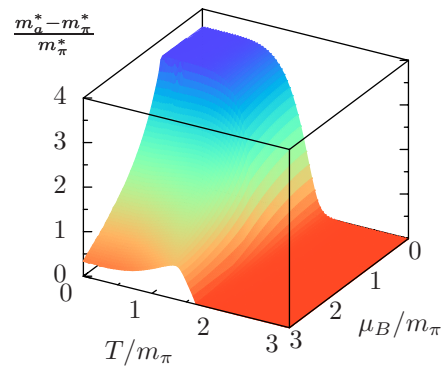}
\caption{Difference of the screening masses of $a_0$ and $\pi$ as a
  function of $\mu_B/m_\pi$ and $T/m_\pi$ in the extreme case of $\alpha=0$.
  This quantity indicates the progressive $\gr{U(1)_A}$ restoration in the
  hot and/or dense medium.  The masses are denoted by $m^\ast$ with
  asterisk to make clear that these are in-medium quantities and
  different from the vacuum $m_\pi$ which is a fixed parameter of the
  model; see Table~\ref{Tab:parameters}.}
\label{Fig:3Dmass}
\end{figure}

Finally, in Fig.~\ref{Fig:3Dmass}, we plot the difference of screening masses
of $a_0$ and $\pi$ as a function of $\mu_B/m_\pi$ and $T/m_\pi$.  Since these
two modes do not mix even in presence of the diquark condensate, the masses can
be calculated straightforwardly also in the BEC phase using the in-medium
propagators~\eqref{medium_props}.  The choice of the screening masses instead
of the real-part ones here is motivated by the lattice measurement, and also
technically favored: while the real-part masses are obscured by Landau damping
in the diquark condensation phase at $T\neq0$, the screening masses remain well
defined.  The results indicate effective restoration of $\gr{U(1)_A}$ symmetry
at high temperature and/or chemical potential.  While in the $T$-direction the
degeneracy more-or-less copies the restoration of chiral symmetry which is
another source of $\gr{U(1)_A}$ breaking, in the $\mu_B$-direction the
convergence of $m_a^*$ and $m_\pi^\ast$ is slower as a result of additional
$\gr{U(1)_A}$ breaking by the diquark condensate.  We note that
Fig.~\ref{Fig:3Dmass} is the extreme example of $\alpha=0$, which is not likely
near the phase boundaries, but could be the case in the quark-gluon plasma
region in view of the lattice data of the topological susceptibility
\cite{Alles:2006ea}. In the future, by combining the two-color lattice outputs
and the PNJL model analysis, it would be possible to make a 3D plot of $\alpha$
which should approach zero at high $T$ and/or high $\mu_B$.


\section{Conclusions}
\label{Sec:conclusions}

We have adopted the PNJL model as an effective approach to two-color
QCD and applied it to the case of two light quark flavors.  This is
the simplest case which exhibits nontrivial low-energy spectrum due to
spontaneous chiral symmetry breaking, and at the same time can be
simulated by lattice Monte-Carlo techniques.  We argued that one can
fit the parameters of the NJL part of the model using physical
(three-color) observables, but their values have to be rescaled
appropriately.  Once this is done and the quark sector is coupled to
the Polyakov loop, the model yields locking behavior of chiral and
deconfinement crossovers as long as $\mu_B$ is zero.

We checked older results on the phase structure of the two-color NJL
model in the plane of $T/m_\pi$ and $\mu_B/m_\pi$, and analyzed its
modification induced by the coupling to the Polyakov loop. The phase
transition between the normal and superfluid phases is second order
for all values of the chemical potential considered. In a large part
of the diquark condensation phase the expectation value of the
Polyakov loop is small, which resembles quarkyonic matter predicted
using large-$N_c$ arguments. We carefully clarified the realization
of quarkyonic matter in the two-color system. The baryon number
density $n_B$ appears finite as soon as $\mu_B$ exceeds the mass of the
baryonic pion, but still the quark contribution to $n_B$ is not
substantial until $\mu_B$ surpasses the twice of the dynamical quark
mass. After then quark degrees of freedom supersede baryons, meaning
a transition from ``superfluid nuclear matter'' into ``quarkyonic
superfluid.''

Our model analysis of the phase diagram is based on two simplifying
assumptions. The first one is the mean-field approximation which treats the
system as a gas of noninteracting fermionic quasiparticles. This may not be
quantitatively accurate in some regions of the phase diagram such as for
$\mu_B\simeq m_\pi$ at nonzero temperature where the system behaves rather as
a dilute Bose gas. On the other hand, in the theory of strongly-interacting
Fermi gases the mean-field approximation is known to be reliable at zero
temperature. Moreover, the structure of the phase diagram concerning diquark
condensation and chiral symmetry restoration is robust, being a direct
consequence of the symmetry of two-color QCD. The second assumption is the
extrapolation of the gauge part of the thermodynamic potential to nonzero baryon
chemical potential. This has been justified for three-color QCD and low baryon
chemical potential by a comparison with available lattice data and, in fact, is
the source of the predictive power of the PNJL model. Therefore, we are
confident about the existence of the quarkyonic superfluid phase as depicted in
Fig.~\ref{fig:phase}. On the other hand, the most recent lattice data
\cite{Hands:2008ha} suggest that the Polyakov loop at a fixed low value of
temperature starts to rise at $\mu_B\gtrsim3m_\pi$, signalling possible
deconfinement. This certainly presents a challenge to the PNJL model and
determines the direction of our future research efforts.

Finally, we studied the dependence of the spectrum of collective
excitations (scalar and pseudoscalar mesons and diquarks) on the
strength of the axial anomaly.  For that sake we introduced a NJL-type
interaction with a tunable $\gr{U(1)_A}$-breaking parameter.  For all
modes we calculated both the real-part mass and the screening mass, which
governs the decay of spatial correlators, in order to facilitate a
direct comparison with lattice simulations.  Above the chiral
restoration/deconfinement temperature the masses of the chiral
partners become degenerate, as expected.  At the same time, our
results indicate that the restoration of $\gr{U(1)_A}$ symmetry in
terms of the masses of parity partners cannot be hidden by chiral
restoration unlike the full topological susceptibility, which is good.
This would naturally incorporate in the model study the suppression of
instanton effects in a hot and/or dense matter.  To make this whole
argument into a quantitative predictive framework, further physical
input from the lattice simulation would be indispensable, e.g.~precise
measurement of $m_a$ as a function of $T$ and $\mu_B$.  It would be
also interesting that, since the topological susceptibility in the
two-color pure gauge theory does not cost much, we can compare the
inferred $\alpha$ behavior and the suppression of the pure topological
susceptibility.

\begin{acknowledgments}
The authors are grateful to H.~Abuki, J.~O.~Andersen, X.~Huang, L.~Kyllingstad,
and W.~Weise for useful discussions, and to L.~He for pointing out an error in
an earlier version of the manuscript. The work of T.~B.\ was supported in part
by the Alexander von Humboldt Foundation, and by the ExtreMe Matter Institute
EMMI in the framework of the Helmholtz Alliance Program of the Helmholtz
Association (HA216/EMMI). K.~F.\ is supported by Japanese MEXT grant No.\
20740134 and also supported in part by Yukawa International Program for Quark
Hadron Sciences. Y.~H.\ was supported by the Grant-in-Aid for the Global COE
Program ``The Next Generation of Physics, Spun from Universality and
Emergence'' from the MEXT of Japan.
\end{acknowledgments}

\appendix

\begin{figure}
\includegraphics[scale=1]{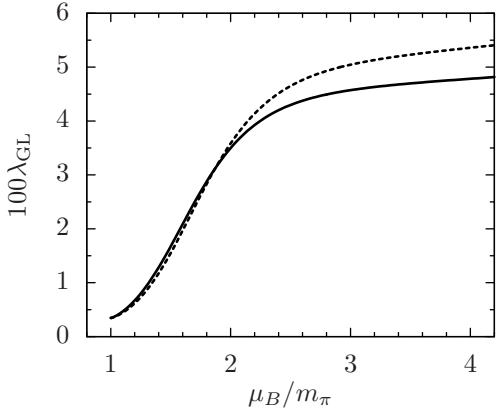}
\caption{Ginzburg--Landau quartic coupling along the second-order
  transition line as a function of $\mu_B/m_\pi$.  The solid and
  dashed lines are the PNJL and NJL model results, respectively, with
  the same input parameters.}
\label{Fig:GLcoupling}
\end{figure}

\section*{Appendix: Ginzburg--Landau expansion of the thermodynamic potential}
\label{app:GL}
In order to make clear whether the phase transition of diquark superfluidity
becomes (weakly) first order at high $\mu_B$, we performed the Ginzburg--Landau
expansion of the thermodynamic potential near the second-order transition line
and calculated the coefficient of the quartic term with respect to $\Delta$.  In
the presence of a tricritical point and the onset of a first-order phase
transition, this coefficient would go to zero and then change its sign to
negative.

The thermodynamic potential $\Omega$ depends only on the square of the
diquark condensate.  In general, $\Omega$ depends on other
condensates, symbolically denoted by $\chi_a$, as well.  In the PNJL
model, we have $\chi=\{\sigma,\Phi\}$, while in the standard NJL model
the only other condensate would be $\sigma$.  To study the behavior of
the diquark condensate near the critical temperature, it is most
convenient to solve the gap equations for the other condensates, i.e.,
$\partial\Omega/\partial\chi_a=0$.  These define $\chi_a$ implicitly
as a function of $\Delta^2$, and the thermodynamic potential is then a
function of $\Delta^2$ solely, that is,
$\Omega=\Omega\bigl(\Delta^2,\chi_a(\Delta^2)\bigr)$.

The coefficients of the quadratic and quartic terms in the
Ginzburg--Landau functional are now determined by the first and second
\emph{total} derivatives with respect to $\Delta^2$, evaluated at
$\Delta=0$.  The first derivative vanishes at the transition point by
means of the gap equation.  The second derivative defines the
effective Ginzburg--Landau quartic coupling and is in general
expressed as
\begin{equation}
 \begin{split}
 \lambda_{\text{GL}} &=
 \frac{d^2\Omega}{d(\Delta^2)^2} \\
 &= \frac{\partial^2\Omega}{\partial(\Delta^2)^2 }
  -\frac{\partial^2\Omega}{\partial\Delta^2\partial\chi_a}
  \!\left(\frac{\partial^2\Omega}{\partial\chi_a\partial\chi_b}\right)^{-1}
  \!\!\frac{\partial^2\Omega}{\partial\chi_b\partial\Delta^2} \,.
 \end{split}
\end{equation}
The inverse in the second term is assumed in the matrix sense.  The
sign of $\lambda_{\text{GL}}$ decides whether the transition is of
first or second order.  In particular in the NJL model, this
expression acquires a simple form (with an obvious notation for the
partial derivatives)
\begin{equation}
 \lambda_{\text{GL}}^{\text{NJL}}=\partial_{\Delta^2\Delta^2}\Omega
 -\frac{(\partial_{\Delta^2\sigma}\Omega)^2}{\partial_{\sigma\sigma}\Omega}\,.
\end{equation}
In the PNJL model, one has to calculate the $3\times3$ matrix of
second partial derivatives.  Given the formula for the thermodynamic
potential~\eqref{TDpot}, these are easily evaluated explicitly at
$\Delta=0$ as,
\begin{align}
\partial_{\Delta^2\Delta^2}\Omega=&\sum_{i=\pm}\int\frac{d^3\vek k}{(2\pi)^3}
\frac{\varphi(\xi^i_{\vek k})-\xi^i_{\vek k}\varphi'(\xi^i_{\vek k})}
{(\xi^i_{\vek k})^3} \,,\notag\\
\partial_{\Delta^2\sigma}\Omega=&2M\sum_{i=\pm}\int\frac{d^3\vek k}{(2\pi)^3}
\frac{\varphi(\xi^i_{\vek k})-\xi^i_{\vek k}\varphi'(\xi^i_{\vek k})}
{\epsilon_{\vek k}(\xi^i_{\vek k})^2} \,,\notag\\
\partial_{\Delta^2\Phi}\Omega=&2\sum_{i=\pm}\int\frac{d^3\vek k}{(2\pi)^3}
\frac{\sinh\beta\xi^i_{\vek k}}{\xi^i_{\vek k}(\cosh\beta\xi^i_{\vek
k}+\Phi)^2} \,,\notag\\
\partial_{\sigma\sigma}\Omega=&\frac1{2G}\frac{m_0}M+4M^2
\sum_{i=\pm}\int\frac{d^3\vek k}{(2\pi)^3}
\frac{\varphi(\xi^i_{\vek k})-\epsilon_{\vek k}\varphi'(\xi^i_{\vek k})}
{\epsilon_{\vek k}^3} \,,\notag\\
\partial_{\sigma\Phi}\Omega=&4M\sum_{i=\pm}\int\frac{d^3\vek k}{(2\pi)^3}
\frac{\sinh\beta\xi^i_{\vek k}}{\epsilon_{\vek k}(\cosh\beta\xi^i_{\vek
k}+\Phi)^2} \,,\notag\\
\partial_{\Phi\Phi}\Omega=&2bT\left[\frac{1+\Phi^2}{(1-\Phi^2)^2}
-24e^{-\beta a}\right] \notag\\
&+4T\sum_{i=\pm}\int\frac{d^3\vek k}{(2\pi)^3}
\frac1{(\cosh\beta\xi^i_{\vek k}+\Phi)^2} \,.
\end{align}
The NJL limit is recovered by setting $\Phi\to1-$.

We plot the numerical results in Fig.~\ref{Fig:GLcoupling}, from which
we conclude that the Ginzburg--Landau coupling is always positive, and
that the phase transition hence is always second order, within a
reasonable chemical potential range (below the cutoff $\Lambda$), and
in both NJL and PNJL models.

\end{document}